\documentclass[]{article}
\usepackage[]{geometry}
\usepackage{graphicx}
\usepackage{epsfig}
\usepackage{amssymb,amsfonts,amsmath,amsthm}
\usepackage{mathrsfs}
\usepackage{epstopdf}
\usepackage{hyperref,color}
\usepackage{natbib}
\usepackage{setspace}
\usepackage{enumitem}

\pdfoutput=1

\begin{document}

\title{One time, two times, or no time?}
\author{Christian W\"uthrich\thanks{This work was supported financially by the John Templeton Foundation (the views expressed are those of the authors not necessarily those of the sponsors). Thanks to the audience at the conference `Einstein-Bergson 100 years later: What is time?', held at the University of L'Aquila on 4-6 April 2019. I warmly thank the organisers of this conference, Eugenio Coccia, Simone Gozzano, Rocco Ronchi, and last but not least Alessandra Campo, for their kind invitation.}}
\date{8 September 2020}
\maketitle

\begin{center}
{\small Forthcoming in Simone Gozzano, Eugenio Coccia, Rocco Ronchi, Alessandra Campo (eds.), {\em Einstein and Bergson 100 Years Later: What is Time?}, de Gruyter.}
\end{center}

\begin{center}
{\em I dedicate this paper to the people of L'Aquila, who had to suffer through the catastrophic earthquake that hit their beautiful town in the early hours of 6 April 2009. 

May The Eagle rise again.}

\end{center}

\begin{abstract}\noindent
Contemporary research programs in fundamental physics appear to suggest that there could be two (physical) times---or none at all. This essay articulates these possibilities in the context of quantum gravity, and in particular of cosmological models developed in an approach called `loop quantum gravity', and explains how they could nevertheless underwrite our manifestly temporal world. A proper interpretation of these models requires a negotiation of an atemporal and a temporal sense of the emergence of (space)time. 
\end{abstract}

\section{Einstein, Bergson, and quantum gravity}
\label{sec:intro}


The date of 6 April is of course noted for another event connected to this conference: the 1922 Einstein-Bergson debate on the occasion of Einstein's visit to the {\em Soci\'et\'e fran\c caise de Philosophie} in Paris. In his brief reply to Bergson, Einstein famously asserted that there was no `philosopher's time', which he took to unduly reify aspects of our experienced time, i.e., of the `psychologist's time'. The trouble with the philosopher's time was, for Einstein, that it contradicted the `physicist's time': the philosopher's time hypostatises the apparent simultaneity of distant events, while the physicist's time, following the insights of special relativity, denies that these events are connected by an objective and absolute relation of simultaneity. At least as expressed in these brief remarks, Einstein accepts the existence of (only) two times: that of the physicist and that of the psychologist. 

The reconciliation of what Wilfrid \citet{sellars1962} called the ``scientific image'', which would include Einstein's physicist's time, with the ``manifest image'' and its psychologist's time constitutes one of philosophy's noblest--- and most urgent---tasks. Einstein's point---as I read him---is that such a reconciliation does not require a {\em tertium quid} in the form of an extraneous structure to be added to those required by the physics and the cognitive science involved in the phenomenology of temporality. Any such addition would be explanatorily idle at best and incoherent at worst. I tend to agree. 

The debate between Einstein and Bergson turns on the then recently discovered relativity of simultaneity, a core element of Einstein's theory of special relativity of 1905. With special relativity, physics had provided deep insights into the elusive nature of time, as it did on numerous other occasions. Our best current understanding of spacetime, and hence of (the physicist's) time, derives from general relativity, a very successful theory of gravity and spacetime formulated by Einstein in 1915. Since general relativity presupposes that the matter which interacts with spacetime is correctly described by pre-quantum physics, however, it will eventually have to be corrected and replaced by a quantum theory of gravity. As quantum gravity remains beyond our empirical reach, physicists have developed diverging attempts to formulate such a theory. Despite their significant differences, many approaches suggest that space and time will not be part of the fundamental furniture of the world, adumbrating the most radical revolution in our understanding of time yet. Instead of being fundamental, space and time are emergent properties of the fundamentally non-spatiotemporal structures postulated by quantum gravity, very much like the solidity of a glass and the liquidity of the water it contains. Just as not any fusion whatever of silicon dioxide combines to form glass or not any hydrogen dioxide forms liquid water, the fundamental degrees of freedom may fail to coalesce into a spatiotemporal form, and thus will not give rise to anything like the space and time we know and love. Only a serendipitous collective action of the fundamental degrees of freedom will form drops---or indeed an ocean---of spacetime and thus deliver a world remotely like ours.

\label{p:mixture}
If borne out, this leads to three scenarios. First, it may or, second, may not be that time emerges from the fundamental structure. The third---tantalizing, but perhaps rather common---possibility is that a world contains both domains with and without emergent spacetime, rendering spacetime a `regional' option. In fact, our universe seems to be just like that: apart from the well-behaved spatiotemporal `phase' we inhabit, it contains a very `early' `epoch', which should not be expected to be spatiotemporal---the big bang. Quantum gravity suggests that `at' or `near' the big bang, we find a non-spatiotemporal `phase'. If this is right, then it would suggest another way in which time might emerge: as a `transition' from an `earlier' non-spatiotemporal to a `later' spatiotemporal states of the universe. How should we categorize the emergence of spacetime---or failure thereof---in these contexts? How can we conceive of a `transition' from timelessness to the regular temporal evolution of our world? Answering these questions will be necessary for an appreciation of the viability of the emergence of spacetime and thus of a quantum theory of gravity, particularly in a cosmological setting. And it constitutes a central piece of a reconciliation of the scientific with the manifest image. The aim of this essay is to stake out some first steps towards this goal.\footnote{I apologise for the many scare quotes, but these are intended to serve as reminders that spatial and temporal locutions are not well-defined in this context.}

This paper offers an accessible presentation and elaboration of some central theses and arguments articulated in \citet{hugwut18}. It presupposes no familiarity with either quantum gravity or the method of contemporary philosophy of physics. It starts in \S\ref{sec:notime} by explicating a remarkable apparent consequence of attempts to bring together quantum physics with general relativity already mentioned above: the disappearance of spacetime from the ontology of these putative fundamental theories. Given that our world is manifestly temporal, time, or something very much like it, must arise from these non-spatiotemporal structures in order for these theories to be empirically coherent and empirically adequate. Time, or, more precisely, spacetime, emerges from the fundamental structures. Models in quantum cosmology in particular suggest that this emergence must itself have temporal aspects. I will exemplify this using cosmological models of loop quantum gravity (LQG). Thus, it appears as if there are two kinds of time present in the objective structure of the physical world. \S\ref{sec:twotimes} articulates these two senses and clarifies their relation. Understanding both senses of time and their relation enables us to arrive at a coherent interpretation of these cosmological models of LQG, suggesting that these may describe either a contracting and re-expanding universe or, more likely, the twin birth of two parallel universes connected at the big bang. These interpretations are considered in \S\ref{sec:onetime}. Brief conclusions and an outlook follow in \S\ref{sec:conc}.

\section{No time? The disappearance of spacetime in quantum gravity}
\label{sec:notime}

Space and time have long been used as convenient expedients to set apart material, physical existence from other forms of existence such as abstract, mental, or divine. According to this venerable standard of partitioning existence, all and only occupants of space and time enjoy {\em physical} existence. As Larry \citet[45]{skl83} exclaimed, a non-temporal, non-spatial world seems ``devoid of real being altogether''. This is natural enough, as our physical world is manifestly spatiotemporal, endowed with a time---with {\em one} time. 

It is precisely this commonplace which gets challenges in contemporary fundamental physics. General relativity (GR) is one of the most successful theories in the history of physics, apparently correctly predicting phenomena such as the deflection of light near massive stars, gravitational time dilation, and the existence of black holes and gravitational waves, to name just a few highlights. It furnishes the basis of our vastly improved understanding of our cosmos and its origin and enables technology such as the global positioning system. GR interprets gravitation not as a force as did Newton's theory, but instead as encoded in the geometry of spacetime---a `geometrization'---, the fusion of space and time necessitated already by special relativity. More particularly, the strength of the gravitational field is given by the amount of curvature in the spacetime continuum, bending the worldlines of objects from stars and planets to humans and zebras to elementary particles in outer space as if the objects attracted one another. To date, no experiment or observation which directly contradicts GR is known. 

Despite its impressive palmar\`es, GR cannot stand as the last word on gravitation and thus on the structure of spacetime. The reason is as simple as it is damning: GR assumes that matter is classical, i.e., as described by classical physics. But if the physics of the twentieth century has taught us anything, then it is that matter cannot be so described: it is irreducibly quantum. In this sense, there is a plethora of experimental evidence invalidating GR, from simple double-slit experiments to the Large Hadron Collider. Just as the location services in every smartphone rely on the insights from GR, its technology is built on the recognition of the deeply non-classical properties of its matter. 

In light of this, GR will have to be replaced by a quantum theory of gravity, by which I designate any theory that can combine the quantum effects of matter with the presence of (strong) gravitational fields. Such a theory is yet to be fully articulated, let alone empirically confirmed. Currently many approaches compete for attention, young talent, and funding. Among them, string theory is most prominent and presently the frontrunner in all three categories, but there is also loop quantum gravity (LQG), non-commutative geometry, causal dynamical triangulation, causal set theory, asymptotic freedom, and many more. The foundational physical principles, the structures or ontologies, even the methods and ambitions differ widely across the field. 

Notwithstanding these differences between research programs, there is a recurring theme in the field: many approaches to quantum gravity either presuppose or entail that fundamentally, there is no space or no time or neither of the two. This denial of space and time takes different forms in different programs, and certainly comes in different degrees. Regardless of these variations, the constancy with which the theme recurs is remarkable. 

This is the sense in which there may be no time.

Before we study a concrete instance of the disappearance of spacetime, let me remark on two generic philosophical challenges a physical theory without space and time in its ontology faces. First, an epistemological point. It may appear as if a non-spatiotemporal theory in this sense is empirically incoherent.\footnote{\citet{hugwut13b} applied the concept of empirical incoherence to quantum gravity. See \citet{yates2020} for a clear recent presentation.} A theory of physics is {\em empirically incoherent} just in case its truth undermines the empirical justification for believing it to be true. A physical theory could thus turn out to be empirically incoherent because it denies a necessary condition of empirical confirmation. A theory {\em sans} spacetime would thus be empirically incoherent since there could not be, it seems, `local beables' which manifest themselves in localized observables in space and time. Space and time and thus the possibility for material objects to occupy, or be located in, space and time appears to be a precondition for making and reporting observations. With space and time inexistent, and thus this possibility denied, it appears impossible to make the observations needed to confirm the theory at stake and thus for justifying belief in it. Empirical incoherence is not logical inconsistency: an empirically incoherent theory is certainly a logical possibility and potentially even a nomological one. There is of course no guarantee that nature is so kind to us as to yield to scientific investigation; but empirical coherence is undoubtedly a presupposition of scientific enquiry and its failure would be devastating to our attempts to discern nature's deep structure. So a physical theory better either be consistent with the fundamental existence of spacetime or else identify the error of this argument.

Second, a more metaphysical concern. By moving to a more fundamental theory, we may have grown to expect that we thereby advance to the next level---at higher energies, or smaller spatial scales---of unveiling the constitution of (part of) our world. Thus, when we supplant GR with a more fundamental theory of spacetime, we would think that we will learn something about the constitution of spacetime---and perhaps of matter, if the theory is unified enough to also be more fundamental than the standard model of particle physics. But constitution, or indeed mereology, appears to be inherently and ineliminably spatial, or perhaps spatiotemporal.\footnote{See \citet{lebihan2018} for a discussion of this point.} How then could a theory denying the fundamental existence of space or spacetime deliver a more fundamental theory of the {\em constitution} of anything? Clearly, such a theory better either be consistent with the fundamental existence of spacetime or else deliver an account of constitution and mereology which is not implicitly spatial. 

We will return to these two challenges in the next section. 
For the remainder of this section, let us look at loop quantum gravity (LQG) to discover one way in which space and time may be absent in a theory of quantum gravity.\footnote{This material draws on \citet{hugwut}, as well as on \citet[\S2]{hugeal13} for the problem of time and on \citet{wut17a} for the various aspects of the disappearance of spacetime in LQG. These sources go into significantly more technical details, which I try to avoid presently.} LQG starts out from GR as our most successful current theory of gravity (and of spacetime), and attempts to transform it into a theory of quantum gravity by applying a `quantization', i.e., an almost algorithmic procedure which turns a classical theory into a quantum one. There are several such quantization recipes available, and just as with cookbook recipes, none of them guarantees success. LQG uses the so-called `canonical' quantization, a procedure which has successfully been applied in other instances, such as the quantization of electrodynamics. Omitting the technical details, what matters for present purposes is that this recipe requires, as a first step, that GR be recast in a particular form, the so-called `Hamiltonian' form. In this form, the physics at stake is captured in terms of a spatially extended physical system which evolves over time according to dynamical equations. This is perfectly adequate, indeed apt, for many phenomena we may be interested in.

However, it appears inept for relativistic physics in which space and time have been fused into one, disregarding GR's central lesson that there is no  absolute `time' external to spacetime itself in which `space' might evolve. This disregard shows in two ways. First, relativistic spacetimes with an `unusual' topology which does not permit a consistent (however non-unique) split of spacetime into space and time are simply declared unphysical and thus omitted. Second, the dynamical equation of GR, Einstein's field equation, is not formally equivalent to the dynamical equation of the Hamiltonian form. In order to establish this equivalence, additional equations called `constraint equations' must be imposed on the Hamiltonian formulation. Such equations always `constrain' some function to be zero, hence their name. One of the constraint equations deserves our attention: the Hamiltonian constraint equation. This equation is remarkable in that the function constrained to vanish is the so-called `Hamiltonian function', which in Hamiltonian mechanics serves to generate the system's dynamics.\footnote{Via `Hamilton's equations' in the classical theory. In the quantum theory, the Hamiltonian operator results in zero when acting on quantum states. Apologies to the reader for all these homonyms, but at least it should not be too difficult to guess the name of the Irish mathematician who developed the formalism.} 

Since the classical Hamiltonian function is thus constrained to vanish, or the quantum Hamiltonian operator annihilates the physical states of the system, it appears as if there cannot be any non-trivial dynamics. Furthermore, in the equations we find no quantity which seems to correspond to a physical time, no `$t$'. In Hamiltonian GR, and any canonical quantization built on it, all physical quantities are constrained to remain constant over a time that does not appear to exist. This is the `problem of time'.\footnote{See, again, \citet[\S2]{hugeal13} and references therein.} Naturally, this is considered a problem since our world is manifestly imbued with ``blooming, buzzing confusion'', thoroughly awash in constant change which occurs over physically real, measurable---indeed experienced---time. That the resulting quantum theory of gravity appears to deny the fundamental existence of either time or change thus stands in crass tension with our manifest picture of the world. Consequently, even some of the founding fathers of LQG have paid considerable attention to resolving this issue either by sketching how to recover time from the fundamentally non-temporal structure or by rejecting the above rough argument to its unwelcome conclusion.\footnote{Cf., e.g., \citet{rovelli2020},\citet{smolin2013}.} 

LQG remains a work in progress. The canonical quantization program turns out to lead to hopeless technical challenges. The partial glance at the fundamental structures it permits despite its unfinished status unveils the following. The (provisional) `Hilbert space' of the system---roughly, the space of its possible quantum states---admits a basis of states with a natural geometric interpretation. In other words, the theory as it is known to date states that the fundamental structure (the `gravitational field') is generically in a `superposition'---roughly, a state of `simultaneous combination'---of these geometrically interpretable base states. These base states are dubbed `spin network states' and can be represented by abstract labelled graphs (see e.g.\ figure 1 in \citealt{wut17a}). 

Spin network states afford a straightforward geometric interpretation---as they are eigenstates of geometrically interpretable operators defined on this Hilbert space. This natural interpretation sees them as discrete, combinatorial structures consisting of grains of space or spacetime connected by adjacency relations. Let us unpack this a bit. First, the structure is often interpreted in the literature (e.g.\ in \citealt[\S1.2.2, \S6]{rov04}) to be spatial (rather than spatiotemporal) in the sense that it is the structure which one expects to give rise to space (rather than spacetime), connected to the fact that the Hilbert space is not the definite one just yet. The correctness of this interpretation, however, is far from obvious, and depends on one's perspective on the role and fate of time and, relatedly, on how to complete the research program of LQG. I will return to this point below.

Second, the structure is discrete or `chunky', giving us a sense that space(time) consists in inscissible atoms of space or spacetime, rather than in an ever more finely divisible continuum. This granularity of spatial, or spatiotemporal, geometry is a direct consequence of the discreteness of the spectra of the geometric operators. The connections between the atoms of space(time) indicate a form of contiguity or adjacency between those that are so connected. Hence, we find a network of atoms of space(time) joined by a neighbouring relation. 

The third point to note is that the generic state of the fundamental structure is a {\em quantum superposition} of these spin network states. Thus, the structure cannot generically be interpreted to have a particular, determinate geometry in the sense of a determinate number of atoms of determinate sizes with determinate facts of the matter which ones among them are adjacent. Generically, the structures are in some state of {\em combination} of networks with different determinate geometries. The sense in which the generic states does not have a determinate geometry is thus analogous to the sense in which a generic electron does not have a determinate spin in a given direction, or, perhaps more accurately, a generic quantum electromagnetic field does not contain a determinate number of photons with a determinate direction of polarization. 

Finally, to repeat, LQG has not so far delivered a final and complete theory---it remains a research program under construction. The {\em pi\`ece de r\'esistance} turns out to be the Hamiltonian constraint equation. Although known in formal outline, the construction of a concrete Hamiltonian operator, let alone the full solution of the resulting equation prove to present formidable technical and conceptual challenges that have resisted convincing and widely accepted resolution to date. These obstacles have not halted the program; instead, physicists have developed two promising work-arounds, to be discussed in the next section.

\section{Two times? The emergence of spacetime}
\label{sec:twotimes}

The first strategy to circumvent the stumbling block of the Hamiltonian constraint equation is to switch half way through from the canonical quantization recipe to the so-called `covariant' one, which seeks a path integral formulation of the theory. This move is based on the insight that at least for simple quantum systems, the two recipes deliver equivalent quantum theories. It is hoped---though of course not proved---that this will be the case here too. This covariant version of LQG takes spin network states as `initial' and `final' states and seeks to express the dynamical content of the theory in terms of transition amplitudes (and hence probabilities) for such pairs of states. Despite the importance of this approach in LQG, we will not further discuss it here.\footnote{See \citet{rovvid15} for an excellent introduction.}

The second strategy simplifies considerably the systems studied to render the technical difficulties manageable, hoping that the lessons learned from such reduced systems transpose into the more generic context. Loop quantum cosmology (LQC) does just this: prior to quantization, it reduces the space of admissible models by imposing an additional condition.\footnote{See \citet{bojowald2008sciam} and \citet[ch.\ 4]{bojowald2010} for a popular introduction.} This condition demands that the model be precisely spatially isotropic and homogeneous, thus reducing the number of degrees of freedom required to account for the system. 

Requiring complete spatial isotropy and homogeneity is precisely what is done in modern cosmology, and hence LQC is thought to capture the `cosmological' sector of LQG, i.e., those models which may in fact describe the large-scale structure of the universe over the course of its history. The so-called `Cosmological Principle' requires that both the spacetime structure and the matter distribution are spatially isotropic around us, i.e., exhibiting the same properties in all directions around us. This principle is reasonably well confirmed by the distribution of the cosmic microwave background radiation and the large-scale distribution of luminous matter in galaxies. However, isotropy around us does not imply either isotropy around any other point in space or full spatial homogeneity, i.e., the sameness of properties everywhere in space. For this, cosmology invokes a much more speculative (though plausible) principle, the `Copernican Principle', according to which we are not privileged or in any way physically special observers. In other words, if the universe is the same around us, it should be the same around every point in space, and so all observers see a spatially isotropic universe. This, in turn, implies that the universe is spatially homogeneous. It can be shown that the so-called `Friedmann-Lema\^{i}tre-Robertson-Walker (FLRW) spacetimes' are the only spatially isotropic and homogeneous spacetimes in GR. 

LQC thus assumes spatial isotropy and homogeneity. Since it also assumes but a single scalar field (clearly an assumption that will have to be revisited), `space' at a `time' can be described by one single degree of freedom describing the `relative size' of space in dynamical evolution, with the scalar field serving as clock variable. In relativistic cosmology, this quantity is the `scale factor' $a(t)$, which is a function of time $t$ (and so presupposes a `cosmic time' $t$, which is, however, available in FLRW spacetimes). Imposing this symmetry (of isotropy and homogeneity) thus leads to a significant simplification of the system studied and makes the attendant mathematical difficulties more manageable. This simplification occurs at the classical level, i.e., before quantization is attempted.\footnote{Although the order of symmetry reduction and quantization should in principle not matter.} In the full theory, the spin networks are generally large, complicated, and irregular, without recurrence or pattern. In contrast to this, the symmetry-reduced quantum configurations of LQC turn out to be highly regular. Consequently, they can be represented by lattice graphs with straight edges of the same length resulting in surfaces of an area of the square of this base length. 

Again, these surfaces are interpreted to represent the universe spatially, with the scale factor giving its `size'. In the quantum theory, there is an operator corresponding to this scale factor. When applied to the spin network states, one can determine the value the scale factor takes for these states, finding that it can be zero or assume one of a (quasi-)discrete set of values.\footnote{The precise sense of `discreteness' in play here is subtle. For more on this, see \citet[123]{wut06}.} Classically, the FLRW models are `singular', i.e., all worldlines of physical objects are past-finite. This is naturally interpreted as the universe having a finite age. One might think that there is thus a time before which there was no time. However, the past-finiteness of the universe does not imply that there was a first (instantaneous) moment; rather, one should think of the cosmic time $t$ as assuming positive values strictly larger than 0. Just as there is no smallest positive real number, there was no first moment in time in those models. The moment $t=0$ is not part of these models, and so corresponds to no physically existing time. The past-finiteness of all worldlines is a sign that the FLRW spacetimes are singular `there',\footnote{In scare quotes because this is not part of the model and so not a physically existing location in spacetime. The singularity is a `global' property of the spacetime.} i.e., they are not mathematically well-defined. This non-place, or the behaviour of the universe in its infancy right `after it', is called `the big bang'.

Another sign of singular behaviour at the origin is the diverging (scalar) curvature `there'. Tracking the relative size of the universe backward in cosmic time it becomes smaller and smaller. As the curvature scales inversely with the scale factor, it grows beyond any bounds as we approach $t=0$. This divergence at the big bang is often seen as a failure of GR itself, as are similar singularities in relativistic models (such as those found in black holes). Peter \citet[156]{bergmann1980} articulated what I take to be a rather common view among physicists:
\begin{quote}
[Singularities] are intolerable from the point of view of classical field theory because a singular region represents a breakdown of the postulated laws of nature. I think one can turn this argument around and say that a theory that involves singularities and involves them unavoidably, moreover, carries within itself the seeds of its own destruction. 
\end{quote}
Consequently, it is a natural expectation that a more fundamental theory of quantum gravity replacing GR ought to eliminate these singularities. This expectation is borne out in LQC, at least to a significant degree:\footnote{For the caveats, see \citet[\S8.1]{wut06}.} the quantum operator corresponding to the scalar curvature is well-defined at what would classically be the big bang. More specifically, the curvature in the model of LQC increases as we trace it back through smaller and smaller scale factors, becomes rather large and peaks at a small (but non-zero) scale factor, before it decreases again to be zero for the zero-size universe at the quantum big bang (see figure 10 in \citealt{wut06}). The quantum model does not exhibit a curvature singularity.

These are merely kinematic facts and do not rely on any `dynamics'. The `dynamical' Hamiltonian constraint equation---and that was to a significant degree the point of the program of LQC---simplifies significantly and can be solved explicitly. Quantum states of the universe which not only satisfy the kinematical constraints, but also this dynamical one are the truly physical states according to the theory. Since these states are the possible states of a universe throughout its history, giving rise (it is hoped) to something like the four-dimensional cosmological models of GR, these states ought to be considered extended in `time'. 

We will return in \S\ref{sec:onetime} to the question of how to interpret the physics of LQC. For now, we want to note that there now seem to be two senses in which time emerges from the fundamental structure of LQG as suggested by LQC. First, as noted above, there are reasons to believe that the fundamental structure according to LQG is not spatiotemporal. If this is right, then it must be shown that at least sometimes, under serendipitous circumstances, the fundamental degrees of freedom must collectively act such as to bring about, at larger scales, space and time, or at least something sufficiently like them. This is the sense in which there must be an {\em atemporal} form of emergence of spacetime: spacetime ontologically depends on more fundamental, ultimately non-spatiotemporal degrees of freedom. 

This atemporal form of emergence is central in the dissolution of the threat of empirical incoherence as articulated in the previous section. By establishing that spacetime emerges at human scales, empirical evidence can unproblematically come in a spatiotemporal form. The additional demand that the spatiotemporality be fundamental cannot be justified. Just as it is not necessary for measuring the temperature of a gas that this temperature somehow be a fundamental properties of the constituents of the gas, it is in no way required that the spatiotemporality be fundamental for the evidence collected in observations and experiments to be manifestly spatial and temporal. The menace of empirical incoherence is thus averted. As for the concern regarding the constitution of spacetime, I have recommended elsewhere that a overly spatial or spatiotemporal concept of constitution be replaced by a functionalist understanding of spacetime \citep{lamwut18}.

To return to the emergence of spcetime, in the cosmological models of LQC, or at least in more realistic ones with many more degrees of freedom, there arises now a second, {\em temporal}, sense in which spacetime ought to emerge. To obtain a more faithful description of the physics of the earliest universe is, together with a deeper understanding of black holes, among the primary objectives of quantum gravity. Given the strength of the quantum effects in the very early universe, we do not expect the classical description of a smooth spacetime to remain applicable. Rather, the universe was born out of a deeply quantum gravitational {\em Ursuppe}. That this {\em Ursuppe} is not spatiotemporal, or at least not `temporal', is further supported by recent findings \citep{brahma2020} according to which the fundamental structures in the very early universe are `spatial', if anything, rather than spatiotemporal: there appears to be a `signature change' from the usual Lorentzian signature to a Euclidean signature. A `Lorentzian' signature is characteristic of a relativistic spacetime with one dimension of time and 3 (or $n$) dimensions of space, whereas a `Euclidean' signature indicates a purely spatial structure. Thus, time truly disappears around the big bang, and there is no sense in which there exists a connected time through that epoch. Time is simply not part of the {\em Ursuppe}. 

If this is correct, then we cannot hope that the physics at that early stage is orderly and spatiotemporal. This is of course again precisely because quantum gravity is generically non-spatiotemporal. In sum, we should expect that cosmological models based on LQG encompass different `phases': an `earlier', non-spatiotemporal, quantum-gravitational phase, as well as a `later' phase, for which the classical spacetime description offered by GR delivers a valid approximation. Thus, it looks as if these models ought to contain a `process' of emergence, a `transition' from the first phase to the second, which involves the (at least approximate) emergence of spacetime. This is the sense in which there must be a {\em temporal} form of emergence spacetime: spacetime arises as an effect of an earlier state, i.e., it depends, perhaps causally, on `prior' states of affairs. 

Thus, we are faced with two distinct notions of emergence and, correspondingly, two ways in which time comes to be. These two notions answer to two distinct, but equally important, questions. First, how can classical relativistic spacetime (or, ultimately, the space and time of our experience) be grounded in a fundamental structure which is non-spatiotemporal? This is arguably the most urgent philosophical question arising in the context of quantum gravity and is extensively addressed in the literature. I sketched only some central points pertaining to this first question here. 

Our focus here is the second question: how can the (fundamentally non-spatiotemporal) universe `evolve' from a non-spatiotemporal `phase' to a spatiotemporal one; i.e., how can there be such a `process' or, more fundamentally, a `change' with time? In fact, how could we possibly order the phases into a `before' and `after'?\footnote{The same problem arises not only in LQC, but also in string cosmology \citep{vene2004sciam,hugwut18} and in Oriti's `geometrogenesis' \citep{ori18}.}

This is the sense in which our world may contain two `times': the emergent temporal aspect of effective spacetime and the temporal aspect of the emergence of spacetime from the {\em Ursuppe} itself. How can we make sense of this puzzling situation? Let us outline a physical interpretation of the models of LQC.

\section{One time? The twin birth of two universes}
\label{sec:onetime}

It would hardly be coherent if there were indeed two distinct notions of physical time, one atemporally emerging from fundamental physics, and another one temporally emerging from earlier physics. If there were in fact two times, it would be surprising if the time of our experience, which ultimately arises from the fundamental structures on which the physics in our experiential vicinity ontologically depends, and the time of the cosmos, born in the big bang, were to coincide. Let us address this threat of incoherence by considering the physical interpretation of the simplest model of LQC and then by ruminating, somewhat speculatively, on extending these lessons to more realistic models. But first, let us prepare the ground for these discussions.

The propaedeutic remark concerns the standard cosmological model of GR, the FLRW spacetime we have already encountered above, and how to think about it. We have already stated that it contains a singularity, the `big bang', marking a `beginning' of the universe. The way cosmologists interpret cosmic history like this is by starting out---`in thought'---from the present state of the universe, and to calculate things backward in time, reaching ever earlier times. In other words, the direction of the evolution of the universe {\em as we theorize about it} is opposite to the direction of the {\em actual, physical} evolution theorized about. Given that we live today, cosmologists, like Schiller's ``Universalhistoriker'' \citeyearpar[127f]{schiller1789}, have no choice but to infer from what is presently the case to what must have transpired before, at earlier and ever earlier times. There is excellent evidence, indirect though it may be, that the classical relativistic model of cosmology offers a surprisingly accurate description of the cosmos for most of the times it assumes to exist, retraced backward by almost 13.8 billion years. As already stated, the model is past finite and is singular. 

Let us now analyze in more detail the simplest model of LQC, i.e., the one with perfect symmetry.\footnote{Physicists also consider slightly more complicated models with a small amount of anisotropy.} The highly regular lattice graphs of varying size are taken to represent the state of the universe at a cosmic time. The Hamiltonian constraint equation delivers the `dynamics' in the sense that it mandates how a set of them can be knit them together into an ordered sequence. The resulting fabric---the maximal family of compossible `instantaneous' states---represents the LQC-equivalent of a four-dimensional spacetime, i.e., the universe throughout its entire history. Physically, such a family can be thought of as an `evolving quantum geometry'. In this simplest, altogether isotropic case, the family members `vary' in size in the sense that the quantum property corresponding to the scale factor differs. 

For the fully isotropic case, the Hamiltonian constraint equation becomes substantially simpler, and in fact turns into a difference equation, rather than a differential equation, as is the usual case for dynamical equations. This is not so surprising, given that the `time' LQC uses to stitch together the `instantaneous' or `momentary' states is discrete, rather than continuous. Differentiation, i.e., building derivatives, requires a backdrop of continuous variation, and hence of a continuum. For a discrete set, we are left with differences, however small they may be. it should be emphasized that the fact that time is discrete does not imply that there are only finitely many `instants' of time; rather, their number is countably infinite. 

As a curious aside, it should be mentioned that this difference equation does not give a momentary\footnote{As by this point in the essay, the reader should be sufficiently `scared', I now unceremoniously drop the scare quotes around `instantaneous', `momentary', `precede', `before' and similar expressions.} state at a time as a function of another momentary state at another time. Instead, it fixes a momentary state as a function of {\em two} other momentary states at different times one unit of time apart. Thus, in order to fully specify initial conditions, we need to specify all momentary states in a (closed) unit interval. 

If the conditions for an interval are thus given, the simplified Hamiltonian constraint equation determines the states at all other times, and thus the entire time-ordered family of momentary states (barring the vanishing of the coefficients in the equation). The hope voiced above that quantum effects may wash out the dynamical singularity of the classical model is indeed borne out, at least almost.\footnote{See \citet[\S8.1]{wut06} for the exceptions.} If, in Schillerian fashion, we evolve a later interval of `initial conditions' backwards in time using the Hamiltonian constraint equation (which does not care about the temporal direction), we find that the evolution continues beyond what classically was the big bang. Beyond the big bang, there is a mirror universe very much like the one we know and love on `this side' of the big bang. 

What is the physics of this newly found realm beyond the big bang? Although parallel und in many ways similar, the two sides are not exactly identical. In the simple model, the orientation of space turns out to be inverted between the two sides \citep[113ff]{bojowald2010}. In more complicated models, one would in addition expect the probabilistic quantum fluctuations to differ. 

How do the two realms connect with one another at the big bang? In the standard interpretation, offered e.g.\ in \citet{bojowald2008sciam} and \citet[ch.\ 4]{bojowald2010}, the model is one in which a large universe shrinks and ultimately collapses to zero size before it rapidly expands again. From slightly more complicated physics, it is believed that the universe heats up before its collapse, only to cool down again as it re-expands after the big bang. Physical time runs unidirectionally from negative infinity to positive infinity all the way through what classically was the big bang. 

\citet[111-115]{bojowald2010} justifies this interpretation with the presence of an effectively repulsive force arising from the discreteness of time. The basic idea is as follows. Given that the quantum state of the universe is described by a wave function, its total energy is proportional to the frequency of the wave function. A discrete time can, however, only support frequencies larger than its characteristic scale of discreteness, thus effectively capping the frequencies. Consequently, the energy any discrete time interval can contain is bounded from above, although its bound may be very large. As the universe contracts and heats up prior to the big bang, the energy density increases. This increase can only be stopped from surpassing the upper bound set by the discreteness of time, says \citet[113]{bojowald2010}, if the collapse itself is stopped and reversed, thus turning into an expansion. In this manner, the discreteness of time thus acts effectively as a repulsive force counteracting, and, for sufficiently large energy densities, dominating attractive gravity. At least given an early, i.e., pre-big-bang, state of contraction, such a bounce is thus what would be expected if time were discrete. 

It thus appears as if this simple model of LQC can explain the sense in which the state around the big bang was prior to our current cosmic era. It ultimately does so in virtue of there being a well-defined size, which can be varied relative to a scalar quantity into the model, which assumes discrete values and acts as cosmic time. Unless more realistic models are considered and confirm the features of the simple model, it remains unclear to what extent, if any, this smooth evolution ought to be taken seriously. In the hope of obtaining a fuller view of the problem, we thus turn to our speculative ruminations about generalizing these lessons to more realistic models. 

It is obvious that the exact isotropy and the perfect symmetry and regularity of the geometry of the momentary states is not a realistic depiction of our real universe. Spatial isotropy and homogeneity are only approximately valid at very large scales: only starting at the order of 500 million light years or so is the universe to a good approximation homogeneous.\footnote{To put this number in context, the closest star from the Sun, Proxima Centauri, is just over 4 light years away, the centre of the Milky Way about 26,000 light years, and the closest major (spiral) galaxy, Andromeda, about 2.5 million light years.} If the universe were perfectly homogeneous, we would not be here to discuss its origin! A more realistic model will thus have to admit  anisotropies and inhomogeneities and exhibit more irregular geometries. If the larger program of LQG is right, then space and time are not directly implemented at the fundamental level, as we have seen above. Thus, given that around the big bang, we will be in the deep quantum-gravitational regime (the reader is reminded, e.g., of the signature change in that epoch), we cannot expect that a simple scalar field will play the role of cosmic clock, as it does in the simple model of LQC with perfect symmetry. If we extrapolate the physics backwards to earlier and earlier times, at some time very soon after the big bang (usually given as around the `Planck time' $10^{-43}$ seconds after the big bang) we arrive at the deep quantum-gravitational regime, where quantum fluctuations are believed to have been so strong that spacetime in the usual sense evaporates. 

The philosophically acute reader will naturally ask what it could possibly mean that the quantum-gravitational period lasted for something like $10^{-43}$ seconds when there is no physically meaningful notion of time present during that epoch. The short answer is {\em nothing}. In the absence of (fundamental or emergent) time, it is simply meaningless to quantify a duration---including setting that duration to zero.\footnote{As is suggested by \citet[1201f]{hugwut18}.} More importantly for our purposes, in the absence of a time ticking continuously and uninterruptedly through the period, it is not meaningful to speak of a unified physical process running from an earlier (pre-big-bang) state to a later (post-big-bang) one, particularly not if there is a signature change. 

However, there is a way something like this scenario would be the correct interpretation, even though the way of referring to it as a single, unified process is strictly incoherent, just as stating that the quantum-gravitational period lasted $10^{-43}$ seconds. Although durations are meaningless, there is a sense in which that epoch occurred {\em before} our era, i.e., is in our past (rather than our future or in no temporal relation to us). Locally, presently, we do have a physical (space)time with a well-defined temporal direction, owing to the second law of thermodynamics. {\em Our} future is distinct from {\em our} past in myriad ways. And if we extrapolate our local time and its direction beyond the scope of its proper applicability, we recognize that the atemporal phase at the big bang is in our past, before our current era. Even if by itself timeless, it is thus meaningful to say that the big bang is in the past relative to our local determination of the direction of time. Hence, it occurred `before' our era.

This extrapolation of local time and its direction has two immediate consequences relevant for the purposes of this paper. First, it unifies the two times identified in \S\ref{sec:twotimes}. The sense in which the big-bang epoch precedes ours is fully due to an extrapolation of our local, emergent time and its arrow. In other words, the temporal emergence of time derives from the atemporal one. The latter is thus prior to the former. In this way, there is an unambiguous and unproblematic way in which we can order the cosmic epoch into a `before' and an `after'. Even so, it should be clear that there is no process in the strict sense from a non-spatiotemporal phase to a spatiotemporal one. If the sense in which the big-bang epoch, i.e., the quantum-gravitational regime, precedes our era derives from the (atemporal) emergence of spacetime and the extrapolation of thermodynamic processes in our local region, then there is just one `time', not two.

Second, it also implies that the standard interpretation of the bounce may not be completely meaningless. As we already noted, the big bang took place in {\em our} past. But similarly, and completely independently, denizens of the other universe may extrapolate their local time and its direction and determine that there exists a `big crunch' to {\em their} future. They would judge the event not a big bang, but a big crunch, since they would be observing that their universe contracts in their local direction of time, and might consequently await their fate with trepidation. Although there would thus not exist a continuous, unified physical process as usually described in the standard scenario, there would be two universes, one of which is contracting to an ever hotter and denser state, and the other expanding and cooling, {\em in their respective (extrapolated) local directions of time}. Although they would be separated from one another in that there is no shared sense of spacetime between them, they would nevertheless form a single physical world, connected by the quantum-gravitational physics at the big bang. To repeat, these connections would neither be spatial nor temporal, but they would most definitely be physical, thus challenging the venerable standard of physical existence mentioned at the outset. What their exact nature turns out to be is for a fundamental theory of quantum gravity to describe---in our case, for LQG. 

However, the absence of a direct temporal relation through the big bang opens up the possibility of an altogether different interpretation: instead of a big bounce from a contracting to an expanding universe, we witness the birth of twin universes from the same quantum {\em Ursuppe}, which are both expanding in their local futures.\footnote{See \citet[132f]{wut06} for a first articulation of this idea in the context of LQC, which has been suggested to me by Carlo Rovelli. This articulation is deepened in \citet[\S4]{hugwut18}. See \citet[\S4.6]{craigsinclair2012} for another recurrence of this idea in the context of LQC (though their analysis differs from mine). See also \citet{bareal2014} for a version of the same idea (of a ``Janus point'' in their later words) in a different physical context.}$^{,}$\footnote{One might ask why {\em two} universes, why not three, or four, or infinitely many? While these may all be options in other models, the mathematics of the simple LQC model naturally suggests (just) two.} Instead of one single process `through' the big bang era, there are two parallel birthing processes from which a spacetime (each) emerges. Time arises twice over, separately and independently in each of the twin branches, though presumably in each case as the result of the propitious coalescence of a large number of fundamental, but individually non-spatiotemporal degrees of freedom. 

More generally, given that for spacetime to emerge from the fundamental degrees of freedom (or, more neutrally, from the fundamental {\em structure}), the conditions must be just right, as indicated in \S\ref{sec:notime}. Just as H$_2$O molecules must be in the right phase for the water to flow as a liquid, the spin networks must be in a sufficiently `geometric' state for spacetime to emerge \citep{wut17a}. And just as it is possible for water to be (near the phase line) in a state of an inhomogeneous mixture of liquid and gas `pockets', one and the same spin network may contain geometric (and hence spatiotemporal) and non-spatiotemporal `regions'. This was the third scenario described on page 2. 
The result would be distinct, isolated islands of spacetime emerging from an ocean that is the fundamental structure. Generically, and unlike in the scenario of the twin birth sketched above where both branches would be connected with one another through a joint structure in their respective local pasts, it would not be the case that these spacetime pockets stand in any meaningful spatiotemporal relation to one another. 

Whether the big bounce or the twin birth scenario obtains in more realistic models of LQC---and whether the loop program is onto a true quantum theory of gravity for that matter---remains to be seen. What I hope to have shown, however, is that the joint appearance of atemporal and temporal forms of the emergence of spacetime can be reconciled in a coherent interpretation of these models.

\section{Conclusion}
\label{sec:conc}


Leaving aside the psychologist's time, Einstein identified two distinct notions of time in his debate with Bergson: the time of the physicist and the time of the philosopher. Whether or not we follow Einstein in denying the existence of the philosopher's time, how can we be sure that there will be exactly one kind of physical time? Contemporary research into fusing quantum physics with relativity theory into a quantum theory of gravity suggests that there may not be a physical time, at least at the fundamental level of existence. Thus, although Einstein took it for granted that there was a physical time, physics itself may end up eliminating time from its fundamental ontology. Does this mean that there may be not time at all?

 It does. If LQG or a similarly non-spatiotemporal theory turns out to be the correct theory of quantum gravity, then it is physically possible that the fundamental structures do not conspire to form space and time; instead, they form a world devoid of space and time and so very much unlike ours. However, any such theory will have to contain models, which give rise to something like the relativistic spacetimes which accurately describe physical aspects of the actual world. In this sense, any such theory must permit the emergence of (space)time under the appropriate circumstances. In other words, it must bring forth the physicist's time. 

In cosmological models based on LQG and other theories of quantum gravity, it appears as if the classically singular big bang is replaced by a quantum foam, which dissolves time (and perhaps space). Therefore, in those models, there must also be a temporal sense in which time emerges from something that is not (yet) temporal. But if time emerges twice over, how do we not end up with two separate notions of time? There is only one resulting (space)time, so whatever these relations of emergence may be, they cannot lead to distinct times. In fact, the local physical time of our era is grounded in fundamental, intrinsically non-spatiotemporal structure, making the atemporal emergence the primary notion. The way in which time emerges temporally is derivative on our local emergent time. This secured the coherence of physical time in light of its potential dual emergence, but also opened up the possibility of a different interpretation of the big bang as the birth of twin universes, rather than a big bounce. 

Although prima facie more secure than the psychologist's and particularly the philosopher's time, the physicist's time remains elusive, potentially ineffable in principle, and perhaps forever beyond our ken. However, I hope to have sketched how its disappearance from the fundamental ontology and its attendant emergence at different scales is a coherent possibility, even in case various forms of emergence come together in producing time.

\bibliographystyle{plainnat}
\bibliography{biblio}

\end{document}